\documentclass[pdflatex,sn-mathphys-ay]{sn-jnl}

\usepackage{graphicx}%
\usepackage{multirow}%
\usepackage{amsmath,amssymb,amsfonts}%
\usepackage{amsthm}%
\usepackage{mathrsfs}%
\usepackage[title]{appendix}%
\usepackage{xcolor}%
\usepackage{textcomp}%
\usepackage{manyfoot}%
\usepackage{booktabs}%
\usepackage{algorithm}%
\usepackage{algorithmicx}%
\usepackage{algpseudocode}%
\usepackage{listings}%
\usepackage{multirow}%
\graphicspath{{./}{figures/}}

\theoremstyle{thmstyleone}%
\theoremstyle{thmstyletwo}%
\theoremstyle{thmstylethree}%

\newcommand{\tabfnm}[1]{\textsuperscript{#1}}
\newcommand{\kms}{{km\,s$^{-1}$}} 

\begin{document}

\title[Magnetic loops in the solar transition region]{Magnetic loops in the solar transition region}


\author{Zhenghua Huang$^{1,2}$ \email{huangzh@nju.edu.cn}}


\affil[1]{Institute of Science and Technology for Deep Space Exploration, Suzhou Campus, Nanjing University, Suzhou 215163, People's Republic of China}
\affil[2]{State Key Laboratory of Lunar and Planetary Sciences, Macau University of Science and Technology, Macau, People's Republic of China}



\abstract{
Transition region (TR) loops are arcade-like features in the solar transition region, with temperatures roughly between $2\times10^4$\,K and $6\times10^5$\,K.
They are a fundamental building block of TR, which are results of the coupling between the magnetic field and the TR plasma.
Their dynamics is closely related to the transport of energy and mass through the TR.
Studies on this class of loops since the launch of the Interface Region Imaging Spectrograph (IRIS) have revealed that they are distinct from coronal loops.
Observations have revealed that they are associated with many small-scale dynamic phenomena in the TR, from which one can infer the physics behind the energy and mass transfer in a magnetically confined environment at TR temperature.
This review summarises the observational results of TR loops, showing their morphology, dynamics, plasma parameters, their relationship with flux emergence, their heating properties, and their implication in the heating of the solar atmosphere.
This class of magnetic loops is much less well understood than their coronal counterparts.
This review also concludes with several critical questions that need to be answered in the coming era with more advanced observational techniques and more precise and realistic simulations.
}




\maketitle

\section{The solar atmosphere and the transition region}\label{sec1}
The solar atmosphere is the outer layers of the sun, where the photons can escape into outer space in the form of sunlight.
The bottom boundary of the solar atmosphere, i.e., the solar surface, is normally defined as the location at which $1/e$ of the emitted energy of photons at 5000\,\AA\ can escape, namely $\tau_{5000}=1$.
The upper boundary of the solar atmosphere cannot be precisely defined.
It extends into interplanetary space in the form of the solar wind, which fills in the entire heliosphere and sometimes is considered as part of the solar atmosphere.

\par
\begin{figure}[!ht]
\centering
\includegraphics[width=0.9\textwidth]{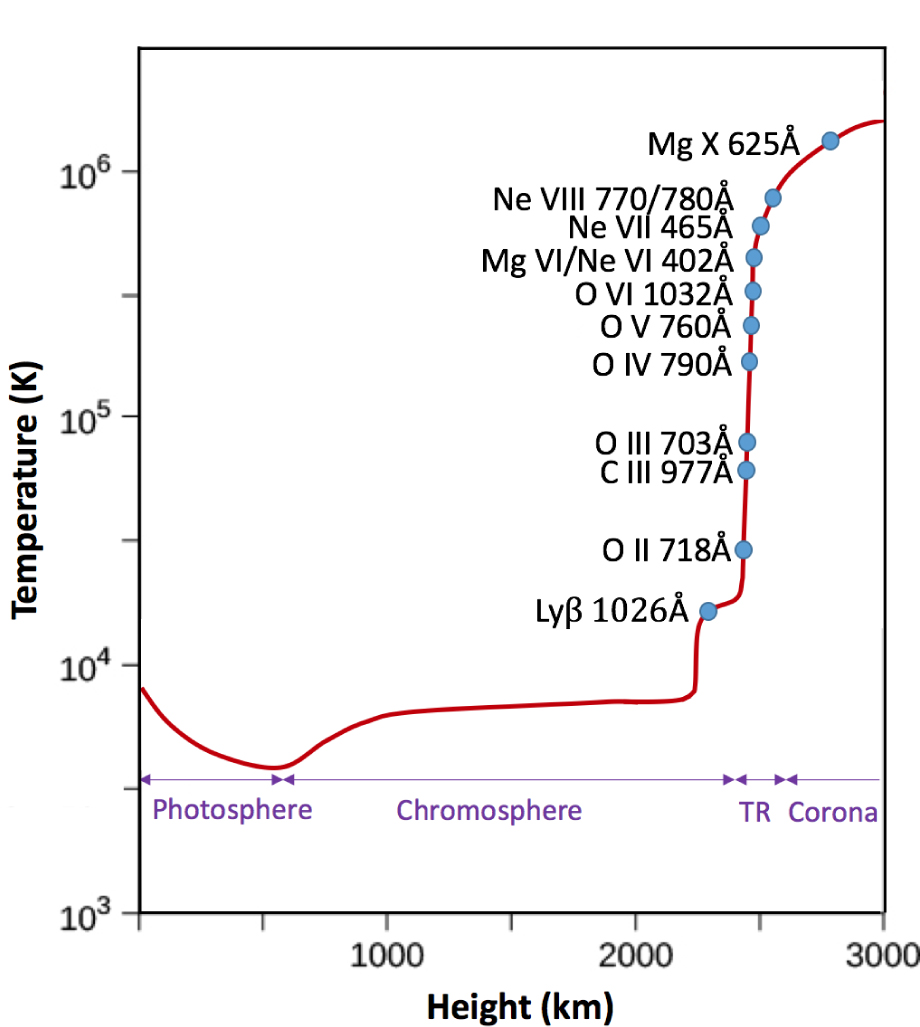}
\caption{Temperature variations along the height above the photosphere as in the one-dimensional model of the solar atmosphere. Typical spectral lines emitted from the TR are denoted on the curve based on their formation temperatures. This image is adapted from the review paper by\,\citet{2017RAA....17..110T}.}
\label{fig01}
\end{figure}

In the classic plane-parallel models\,\citep{1981ApJS...45..635V,2019ARA&A..57..157C}, the stratified solar atmosphere includes three major layers, photosphere, chromosphere, and corona (Fig.\,\ref{fig01}).
The photosphere is the innermost layer and the visible surface of the sun, with a thickness of a few hundred kilometers, yet it emits most of the sunlight.
At the bottom of the photosphere, the temperatures are about 6\,000--7\,000\,K, and at its top, the lowest temperature in the solar atmosphere is about 4\,500\,K.
The chromosphere lies above the photosphere, which has a thickness of about a thousand kilometers.
The temperatures in the chromosphere gradually rise to 10\,000--20\,000\,K at its top.
The corona is the outermost layer that extends to several solar radii.
It is the hottest part of the solar atmosphere, with temperatures around a million Kelvin.
This abrupt temperature increase to a million degrees constitutes the so-called ``coronal heating problem'', which is one of the longstanding questions in solar physics.

\par
Between the chromosphere and corona, the temperature rapidly increases from 20\,000\,K to about 1\,000\,000\,K within just a few tens of kilometres.
This defines a critical interface in the solar atmosphere, namely the transition region (TR).
Through this region, most elements of the sun, including the two most abundant elements (hydrogen and helium), become fully ionised (see Table\,\ref{tab01}).
As a result, the radiation from the TR is dominated by emission lines from various ions.
Collisions (via thermal) are the major ionisation process in the TR\,\citep{1975A&A....44..321N},
and ionisation has a profound effect on the thermal equilibrium and the coupling between the magnetic field and the plasma\,\citep{2019ApJ...872..123B}.
 In other words, a major part of the energy that pours into the TR is used for ionisation.
Because radiation takes only a small fraction of the energy, also the region has a short cooling time scale, any brightenings, as observed by the remote sensing facilities, cannot sustain as long as those in the corona.
This is also a reason why the observed TR is highly dynamic.

\par
\begin{figure}[ht]
\centering
\includegraphics[width=0.9\textwidth]{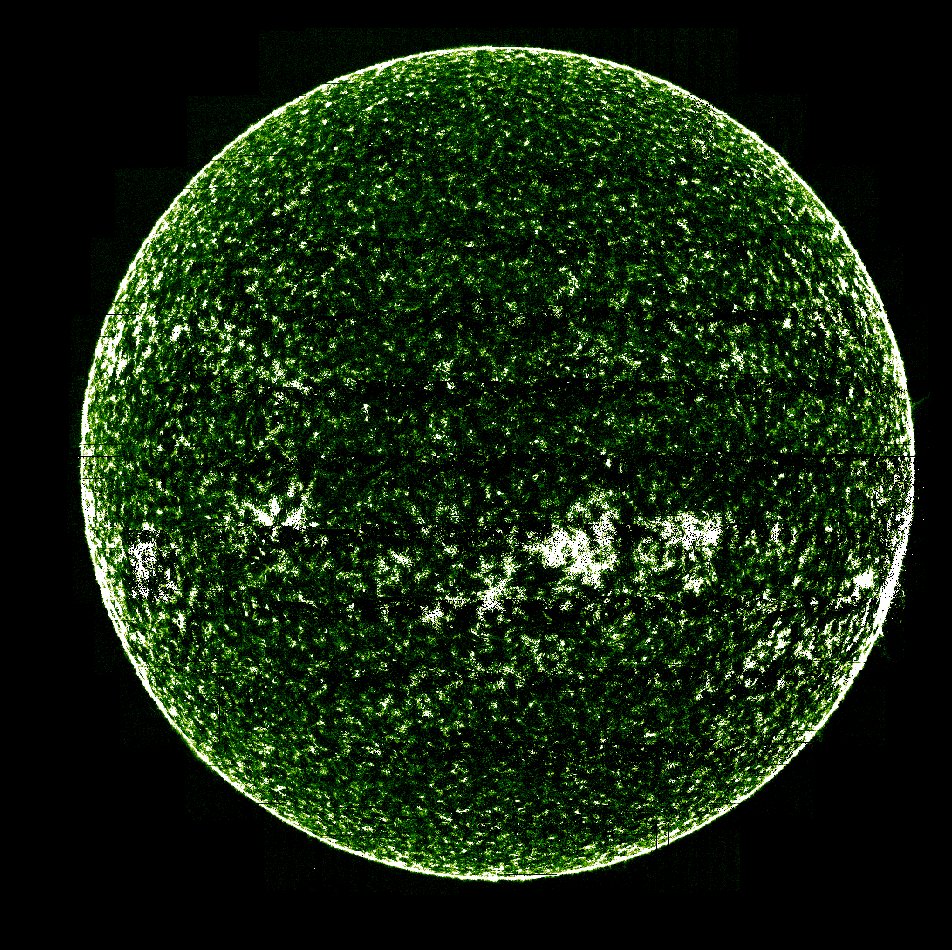}
\caption{Full disk mosaic of the sun on May 27, 2014, as observed in Si\,{\sc iv}~1394\,\AA, which has a formation temperature of $8.0\times10^4$\,K.
This image was combined from tens of raster scans taken by the spectrograph of IRIS.
The image clearly reveals the presence of active regions (large, bright spots toward the southwest near the disk center) and the network (small, bright patches spreading across the disk) at TR temperatures.}
\label{fig02}
\end{figure}

Observations obtained from many generations of instruments, such as the High Resolution Telescope and Spectrograph\,\citep[HRTS,][]{1990ppst.conf...29D} of Naval Research Laboratory (NRL) rocket missions, the Solar Ultraviolet Measurements of Emitted Radiation\,\citep[SUMER,][]{1995SoPh..162..189W} and the Coronal Diagnostic Spectrometer\,\citep[CDS,][]{1995SoPh..162..233H} on board the SOHO mission\,\citep{1995SoPh..162....1D}, the Transition Region and Coronal Explorer\,\citep[TRACE,][]{1999SoPh..187..229H}, the EUV Imaging Spectrometer\,\citep[EIS,][]{2007SoPh..243...19C} for Hinode\,\citep{2007SoPh..243....3K},
the Interface Region Imaging Spectrograph\,\citep[IRIS,][]{2014SoPh..289.2733D} and the Extreme Ultraviolet Imager\,\citep[EUI,][]{2020A&A...642A...8R} aboard Solar Orbiter\,\citep[SolO,][]{2020A&A...642A...1M} have revealed that the TR has an extremely dynamic nature.
Figure\,\ref{fig02} shows a full disk mosaic image of the sun emitting at a temperature of about $8.0\times10^4$\,K,
from which we can clearly see that the TR is highly structured.
The coupling between the ionised plasma and the solar magnetic field gives rise to many magnificent phenomena in the TR.
In this region, many small scale transients are ubiquitous, including small-scale reconnection events\,\citep[e.g.][etc.]{1989SoPh..123...41D, 1997Natur.386..811I, 2014ApJ...797...88H,2018ApJ...856..127G}, intense UV bursts\,\citep[e.g.][etc.]{2014Sci...346C.315P,2015ApJ...811..137T,2015ApJ...812...11V,2016ApJ...829L..30H,2018ApJ...854..174T,2021Symm...13.1390H}, small-scale jets\,\citep[e.g.][etc.]{2014Sci...346A.315T,2016SoPh..291.1129N,2018ApJ...868L..27P,2019SoPh..294...92Q,2019ApJ...873...79C,2021NatAs...5...54A} and dynamic loops\,\citep[e.g.][etc.]{2014Sci...346E.315H,2015ApJ...810...46H,2018ApJ...856..127G,2018ApJ...869..175H,2019ApJ...887..221H,2019ApJ...874...56R}.
These TR variabilities could be impulsive response to coronal heating\,\citep[e.g.][etc.]{2014Sci...346B.315T,2020ApJ...889..124T,2023ApJ...945..143C,2026ApJ..1000...31C} and thus they are a critical aspect of the mass and energy coupling in the entire solar atmosphere.
More about the dynamics of the TR can be found in several recent review articles \citep[e.g.][etc.]{2017RAA....17..110T,2018SSRv..214..120Y,2019STP.....5b..58H,2019ARA&A..57..189C,2021SoPh..296...84D,2023hxga.book..134T}.

\par
In this review, I focus on loop structures in the TR, namely TR loops, whose full lengths have temperatures roughly between $2\times10^4$\,K and $6\times10^5$\,K.
Comparing to their widely-studied coronal counterparts, coronal loops, TR loops have significantly different physical characteristics and dynamic evolutions as revealed by high-resolution observations achieved by several successful solar missions in the past decade.
In what follows, a general introduction to loop in the solar atmosphere is given in Section\,\ref{sec2}; observations to TR loops are summarised in Section\,\ref{sec3}; a summary of existed modelling results of TR loops is provided in Section\,\ref{sec4}; and a summary to this review with a few open questions is given in Section\,\ref{sec5}.

\begin{table}[h]
\caption{Top-10 abundant elements on the sun and their first ionisation energies.}\label{tab01}%
\begin{tabular}{@{}ccccc@{}}
\toprule
\multirow{2}{*}{Elements} & \multirow{2}{*}{Abundances\tabfnm{[a]}} & First ionisation & Equivalent&Max single ionisation\\
 &   & potential\tabfnm{[b]} (eV) & temperature\tabfnm{[c]} (K)& temperature\tabfnm{[d]} (K)\\
\midrule
H    & 12.00 & 13.59  & $1.05\times10^5$ & $7.94\times10^4$ \\
He  & 10.91 & 24.59  & $1.90\times10^5$ & $5.01\times10^4$\\
O    & 8.69   & 13.62  & $1.05\times10^5$ & $2.82\times10^4$\\
C    & 8.46   & 11.26  & $8.70\times10^4$ & $2.52\times10^4$\\
Ne  & 8.06   & 21.56  & $1.67\times10^5$ & $3.55\times10^4$\\
N    & 7.83   & 14.53  & $1.12\times10^5$ & $2.82\times10^4$\\
Mg  & 7.55   & 7.65   & $5.90\times10^4$ & $1.41\times10^4$\\
Si    & 7.51   & 8.15   & $6.30\times10^4$ & $1.58\times10^4$\\
Fe   & 7.46   & 7.90   & $6.10\times10^4$ & $1.41\times10^4$\\
S     & 7.12   & 10.36 & $8.00\times10^4$ & $1.78\times10^4$\\
\botrule
\end{tabular}
\footnotetext{\tabfnm{[a]} The data are from\,\citet{2021A&A...653A.141A}. The abundance of an element $X$ is defined as $A(X) = \log_{10}\left(\frac{N_X}{N_H}\right) + 12$, where $N_X$ is the number density of the element $X$ and $N_H$ is that of the hydrogen.}
\footnotetext{\tabfnm{[b]}Data are from \url{https://physics.nist.gov/PhysRefData/ASD/ionEnergy.html}.}
\footnotetext{\tabfnm{[c]}Temperature of electrons corresponds to an internal energy that is equivalent to the first ionisation potential.}
\footnotetext{\tabfnm{[d]}Temperature at the maximum of the single ionisation of the element obtained from CHIANTI database\,\citep{1997A&AS..125..149D,2021ApJ...909...38D}.}
\end{table}

\begin{figure}[ht]
\centering
\includegraphics[width=0.9\textwidth]{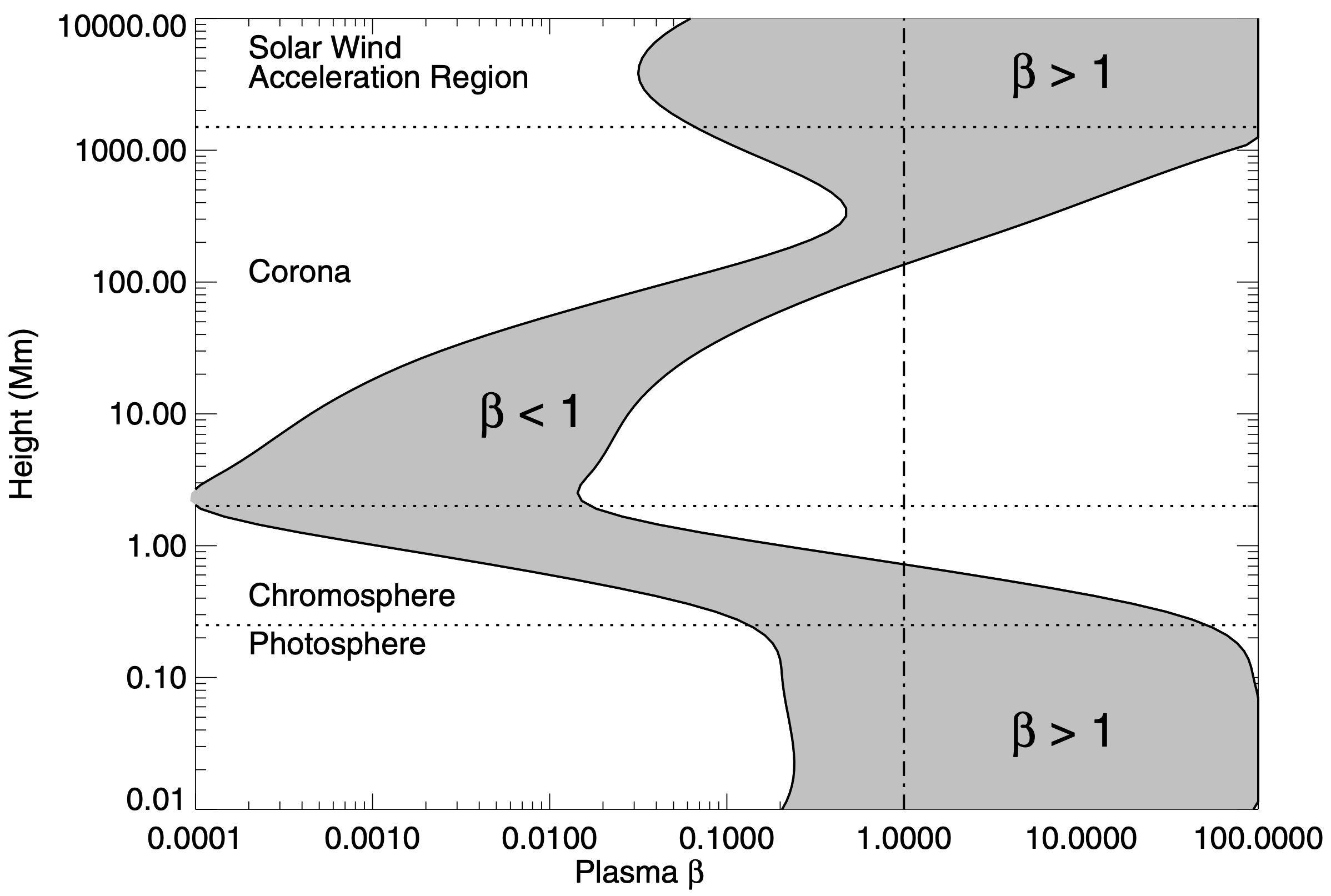}
\caption{Variation of plasma-$\beta$ in the solar atmosphere as a function of height. This image is adapted from the book of ``Physics of the Solar Corona'' of \citet{2005psci.book.....A} that is originally from \citet{2001SoPh..203...71G} with minor modifications.} 
\label{fig03}
\end{figure}

\begin{figure}[!ht]
\centering
\includegraphics[trim=3cm 3cm 3cm 3cm, clip, width=0.9\textwidth]{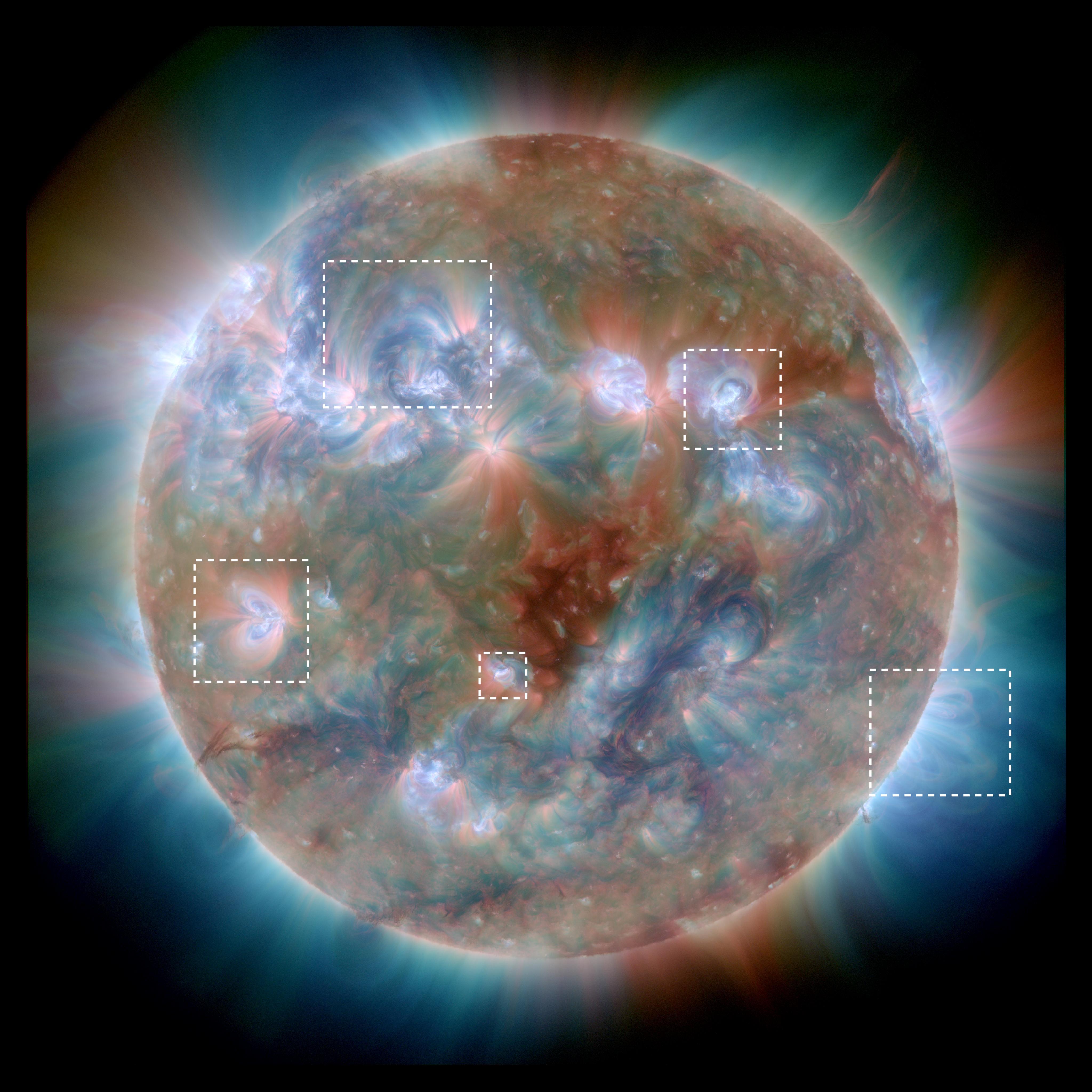}
\caption{Three-colour composition image of the full disk of the sun based on the observations taken on 1 November 2022 around 16:50\,UT by SDO/AIA in the passbands of 171\,\AA\ (red), 193\,\AA\ (green), and 211\,\AA\ (blue).
A few regions denoted by squares are abundant with loops.} 
\label{fig04}
\end{figure}

\section{Loops in the solar atmosphere}\label{sec2}
When the plasma in the solar atmosphere begins to be ionised, its coupling with the solar magnetic field becomes strong, and it starts to distribute along the magnetic field lines.
The plasma-$\beta$ is often used to characterise the interaction between the plasma and magnetic field, which is defined as the ratio of the thermal pressure to the magnetic pressure, and in cgs units, it expresses 
$$\beta=8\pi nk_BT/B^2,$$ 
where $n$ is the number density of the plasma, $k_B$ is the Boltzmann constant and $B$ is the magnetic field strength.
While plasma-$\beta$ is less than 1, which is the case in the upper chromosphere, TR, and the corona (see Fig.\,\ref{fig03}), the magnetic field dominates, and the solar atmosphere is structured by the magnetic field.

\par
A notable structure in the upper solar atmosphere, representative of the coupling between the solar magnetic field and the ionised gas, is the so-called loop, which connects two magnetic elements on the solar surface with opposite polarities.
They are arch-like, bright structures that stand out from the ambient solar atmosphere.
Fig.\,\ref{fig04} shows a tri-color composite image of the full disk of the sun observed by the Atmospheric Imaging Assembly\,\citep[AIA,][]{2012SoPh..275...17L} on board the Solar Dynamics Observatory\,\citep[SDO,][]{2012SoPh..275....3P},
from which we can see that loops are one of the fundamental building blocks of the solar atmosphere.

\par
The lengths of loops vary from a few to hundreds of mega-meters (Mm) \,\citep[see review by ][]{2014LRSP...11....4R}.
Short loops with sizes from a few Mm to about 30 Mm connecting small-scale opposite polarities usually locate in quiet sun regions and form localised brightenings, such as coronal bright points\,\citep[e.g.][]{1981SoPh...69...77H,2012A&A...548A..62H,2016ApJ...818....9M,2023A&A...678A..32M} and even smaller brightenings revealed by SolO/EUI\,\citep{2021A&A...656L...4B,2021A&A...656L...7C,2021A&A...656L..16M}. They are also present in the cores of active regions\,\citep{2018ApJ...869..175H,2019ApJ...887..221H}.
Longer loops with sizes up to hundreds of Mm normally shape the outer boundaries of an active region and connect its two major opposite polarities, or connect opposite polarities in a pair of active regions\,\citep{2014A&ARv..22...78W} that sometimes even locate in opposite hemispheres\,\citep{1977SoPh...52...69S,1996ApJ...456L..63T,2000ApJ...531..553P}.

\par
Because loops are easy to trace in the solar atmosphere and connect different layers and locations, they are ideal objects for studying mass and energy transfer in the solar atmosphere.
Many mysteries of the solar atmosphere have been uncovered by studies of the morphology and dynamics of loops.
For example, the heating mechanism of coronal loops can be inferred from the relations among their temperatures, pressures, and lengths\,\citep{1978ApJ...220..643R,1995Natur.377..131K,2017ApJ...842...38X};
oscillations in loops are a good proxy to probe the magnetic field in the corona\,\citep[see reviews by][]{2005LRSP....2....3N,2012RSPTA.370.3193D,2020ARA&A..58..441N};
and small-scale dynamics in loops reveals braiding reconnection processes in the corona\,\citep{2013Natur.493..501C,2018ApJ...854...80H,2025ApJ...985...17Z,2025ApJ...995...94C}.
Therefore, loops have been studied intensively since the 1960s, and they have helped improve our understanding of the nature of the solar atmosphere\,\citep{2015RSPTA.37340256K,2015RSPTA.37340257S}.

\section{Observations of TR loops}\label{sec3}
Loops have also been thought to be a basic structures of the TR since the 1980s\,\citep{1983ApJ...275..367F,1986ApJ...301..440A,1987ApJ...320..426F,1998ApJ...507..974F,2001ApJ...558..423F,1986SoPh..105...35D,1993ApJ...411..406D,2012A&A...537A.150S}, but solid observational evidence only becomes available after the launch of the Solar and Heliospheric Observatory (SOHO), which are ascribed to raster scans provided by two spectrographs (SUMER and CDS) with spectral lines formed at the TR temperatures\,\citep{1997SoPh..175..511B,1999SoPh..190..379B,2000A&A...357..743L,2000ApJ...533..535C}.
After the launch of IRIS in 2013, the dynamical evolution of TR loops can be followed thanks to the special design of the instrument that allows both imaging and spectral data to be acquired simultaneously at high spatial resolution and temporal cadence\,\citep{2014Sci...346E.315H}.

\par
In this section, the observational aspects of TR loops are summarised.
TR loops rooted in active regions and quiet-sun regions are distinct from each other likely due to the difference in the magnetic dynamics (i.e. local dynamos), and their observational results are summarised separately in Subsections\,\ref{ssec31} and \ref{ssec32}.
While small-scaled activities associated with loops are crucial for understanding how the magnetic field acts in heating and reshaping the solar atmosphere, this aspect related to TR loops is given in Subsection\,\ref{ssec33}.

\begin{figure}[!ht]
\centering
\includegraphics[width=\textwidth]{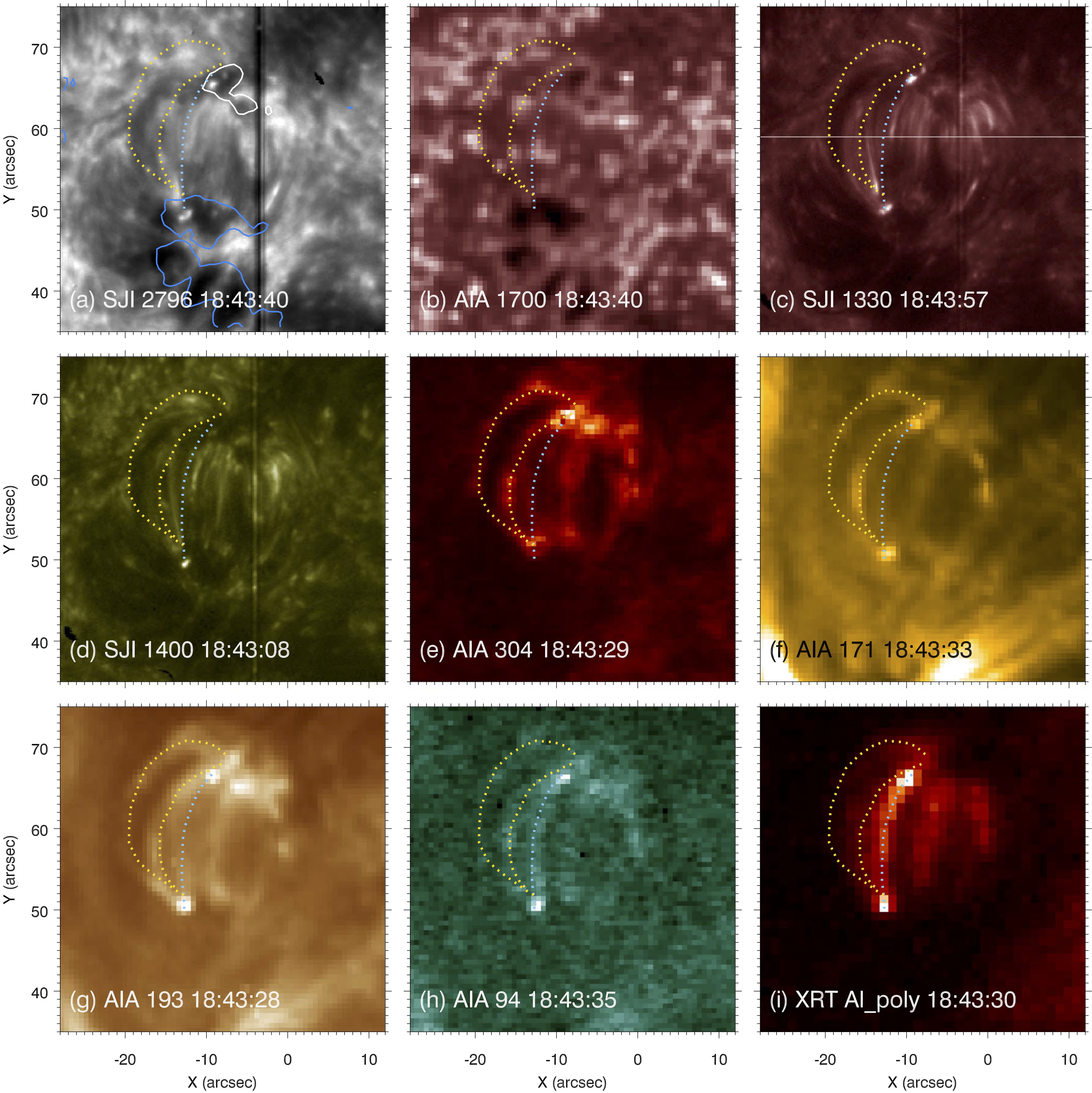}
\caption{Images of a solar region abundant with loops as seen in IRIS SJ 2796\,\AA\ (a), AIA 1700\,\AA\ (b), SJ 1330\,\AA\ (c), SJ 1400\,\AA\ (d), AIA 304\,\AA\ (e), AIA 171\,\AA\ (f), AIA 193\,\AA\ (g), AIA 94\,\AA\ (h) and XRT Al\_poly (i).
The cyan dotted lines denote a loop thread that is clearly seen in the XRT image, and the yellow dotted lines denote two loop threads that are clearly seen in the AIA 304\,\AA\ image.
This figure is reused with permission by \citet{2018ApJ...869..175H}.}
%
\label{fig05}
\end{figure}


\subsection{TR loops in active regions}\label{ssec31}
\begin{figure}[!ht]
\centering
\includegraphics[width=0.9\textwidth]{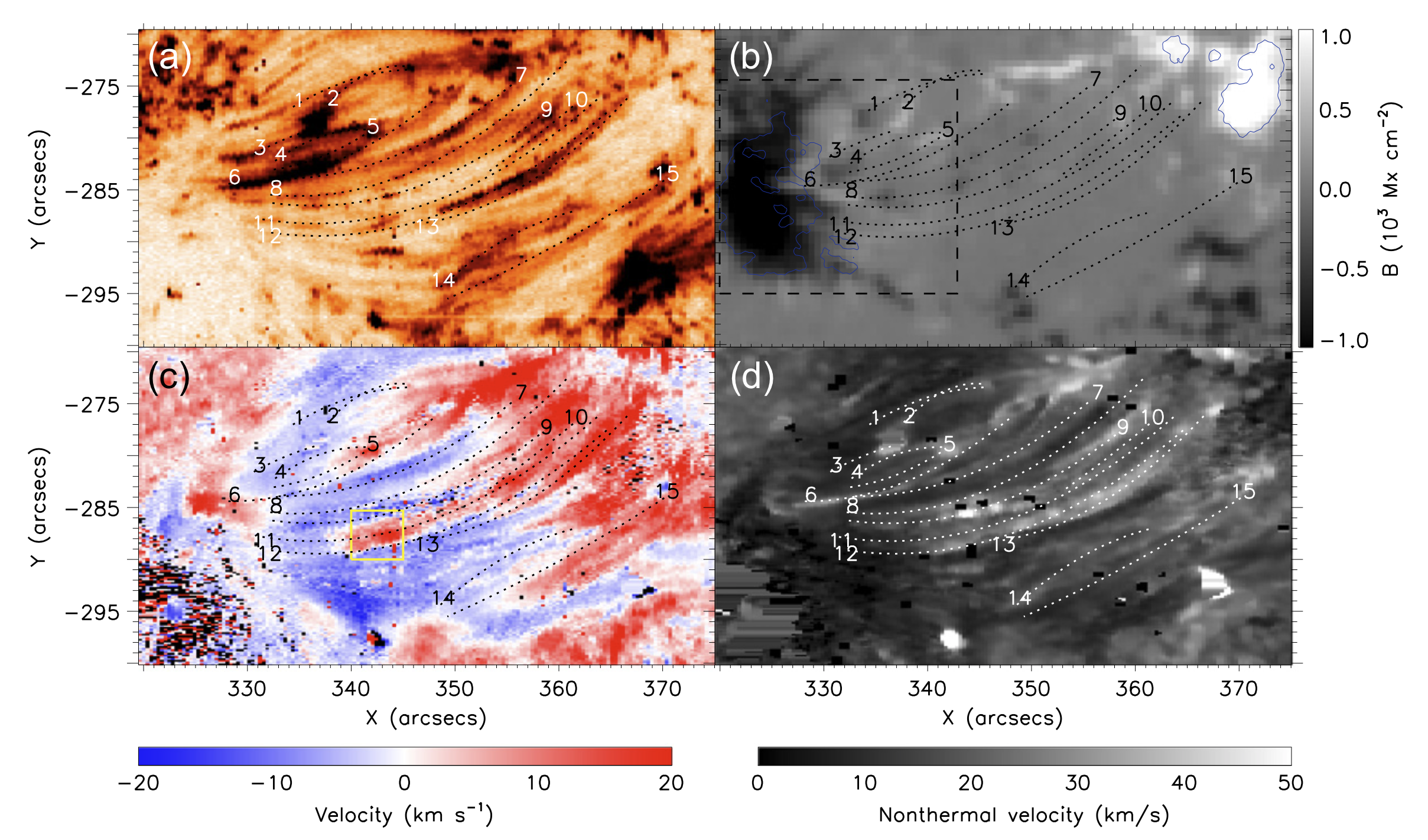}
\caption{Active region observed in Si\,{\sc iv}\,1403\,\AA\ (panels (a), (c)\& (d)) and SDO/HMI magnetogram (panel (b)), which locates near the center of the solar disk.
The Si\,{\sc iv}\,1403\,\AA\ radiance map of the region is shown in panel (a), in which an inverted colour table is used, and 15 TR loops are marked as dotted lines with numbers.
The Si\,{\sc iv}\,1403\,\AA\ Doppler velocity map of the region is shown in panel (c), in which we can see clearly that the Doppler velocities change from blue-shifts in their east footpoints to red-shifts in their west footpoints.
The Si\,{\sc iv}\,1403\,\AA\ non-thermal velocity map of the region is shown in panel (d), in which the TR loops apparently have greater nonthermal motions.
The figure is adapted from \citet{2015ApJ...810...46H}.} 
\label{fig06}
\end{figure}

In active regions, loops with various temperatures are usually found.
A study\,\citep{1997SoPh..175..487F} based on CDS observations of active regions identifies several classes of loops, including multi-thermal loops seen in all temperatures from $2\times10^5$\,K to $2\times10^6$\,K, warm loops seen at $10^6$\,K but not reaching $1.6\times10^6$\,K, cooler loops seen at 2--4$\times10^5$\,K and occasionally reaching $6\times10^5$\,K but not exceeding $10^6$\,K.
Observations from Hi-C rocket mission\,\citep{2014SoPh..289.4393K} also reveal loops with temperatures of $2.5\times10^5$\,K connecting two moss regions\,\citep{2013ApJ...771...21W}.
This demonstrates that loops in TR temperatures are a distinct group of loops in active regions, which are much more dynamic than their coronal counterparts\,\citep{1997SoPh..175..487F,1997SoPh..175..511B,2003A&A...406..323D}.
As shown in Figure\,\ref{fig05}, the loop structures seen in SJ 1400\,\AA\ (TR loops) are significantly distinguished from those hotter ones (e.g. in AIA 193\,\AA, XRT Al\_poly).
TR loops at lower temperature of about $7.9\times10^4$\,K have been discovered by IRIS observations\,\citep{2015ApJ...810...46H}, as shown in Figure\,\ref{fig06}.
Please note that the legs of coronal loops or fan loops often have a response at TR temperature\,\citep{2003A&A...406.1089D,2007PASJ...59S.727Y,2009ApJ...695..642U}. 
``TR loops'' described here refer to those loops that can be discerned in full length at TR temperatures.
Some TR loops might also be temporarily heated to coronal temperatures while strong heatings take place\,\citep{2018A&A...611A..49Y}, but that can only sustain for a short period of time.

\par
Strong flows are one of the dynamic phenomena frequently present in TR loops.
The CDS observations of a loop at about $2\times10^5$\,K has revealed a flow of about 50\,\kms\ along the loop axis\,\citep{1997SoPh..175..511B}, which is also confirmed by SUMER data\,\citep{2000ApJ...533..535C} and Hinode/EIS observations\,\citep{2009ApJ...695..642U}.
Flows at the footpoints of TR loops might show inversion between upflows and downflows at a temperature of about $2\times10^4$\,K\,\citep{2019ApJ...874...56R}.
Clear evidence of siphon flows with speeds of about 10\,\kms\ in TR loops rooted in active region near the solar disk center has been revealed by IRIS observations in Si\,{\sc iv} ($7.9\times10^4$\,K)\,\citep{2015ApJ...810...46H}, as shown in Figure\,\ref{fig06} (c).
Such a siphon flow is clear evidence of an imbalance of heating sources in the footpoints of the TR loops, and enthalpy flows are crucial in distributing heat along the entire loop\,\citep{2006A&A...452.1075D}.

\par
As loops are a good proxy for tracking magnetic geometry in the solar atmosphere, understanding how they form, along with flux emergence, is critical for understanding the formation of the multi-layered solar atmosphere.
Such studies have been carried out by using observations of IRIS, AIA, EIS, XRT, and HMI\,\citep{2018ApJ...869..175H,2019ApJ...887..221H}. 
Some critical results include: (1) emerging loops tend to be more isothermal than multithermal; (2) TR loops are an order of magnitude denser than coronal loops rooted in the same region; (3) TR loops are expanding by contrast to that coronal loops are contracting; (4) interactions among TR loops and/or ambient field can lead to heating therein, but no evidence shows that any TR loop can be directly heated to coronal temperatures; (5) periodic brightenings with a period of 2--3 minutes are found in the emerging loops suggesting that heatings therein might be modulated by oscillations in the photosphere.
These results indicate that TR loops may be a direct product of emerging flux, whereas coronal loops require additional heating.

\begin{figure}[!ht]
\centering
\includegraphics[width=\textwidth]{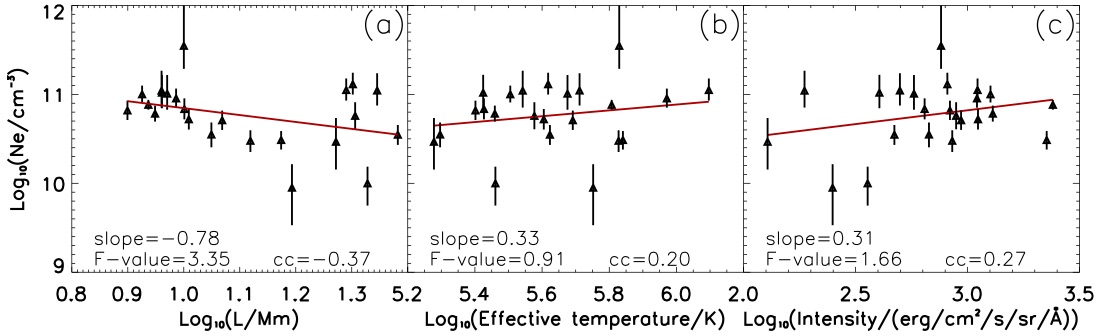}
\caption{Relationships between electron densities and loop lengths ((a)), between electron densities and effective temperatures ((b)) and between electron densities and radiative intensities((c)).
The red solid lines show the linear functions given by the regression analysis to the data points.
The figure is adapted from \citet{2024ApJ...966L...6L}.} 
\label{fig07}
\end{figure}

The sizes of TR loops might vary from one individual to another.
A study of 15 TR loops in steady status based on the IRIS spectral data in Si\,{\sc iv}\,1402.8\,\AA\ found that their apparent lengths vary from 7\,Mm to 30\,Mm\,\citep{2015ApJ...810...46H}. 
Their cross-sections are about 500\,km, but one should keep in mind that this is close to the instrument's resolution limit\,\citep{2015ApJ...810...46H}.
A statistics with 25 samples, considering the extensions of TR loops to magnetic polarities, gives a range of 8.03--29.64\,Mm for their lengths\,\citep{2024ApJ...966L...6L}.

\par
The electron density of a TR loop also varies from case to case and depends on its temperature.
A case study of loop observed in Mg\,{\sc vii} ($T\sim6.3\times10^5$\,K) by Hinode/EIS gives densities of $3\times10^9$\,cm$^{-3}$ at its footpoint and $1.5\times10^9$\,cm$^{-3}$ at a projection height of 20\,Mm\,\citep{2007PASJ...59S.727Y}.
In an emerging TR loop at $8\times10^4$\,K, the electron density is found to be about $5.0\times10^{10}$\,cm$^{-3}$\,\citep{2018ApJ...869..175H}.
In a transient loop at about $1.4\times10^5$\,K, the density is measured as $6.3\times10^{10}$\,cm$^{-3}$\,\citep{2022MNRAS.512.3149G}.
A systematic study of 25 TR loops based on IRIS O\,{\sc iv} (T$\sim1.4\times10^5$\,K) observations\,\citep{2024ApJ...966L...6L} shows that their electron densities range from $8.9\times10^{9}$\,cm$^{-3}$ to $3.5\times10^{11}$\,cm$^{-3}$.

\par
Based on spectral data, effective temperatures of TR loops can be derived from the broadening of spectral lines, which are representative of the thermal or thermal-like velocities of ions at the emission source.
The analysis of 25 TR loops shows that the effective temperatures of the same TR loop observed in Si\,{\sc iv} (T$\sim8\times10^4$\,K) is normally greater than that in O\,{\sc iv} ($T\sim1.4\times10^5$\,K)\,\citep{2024ApJ...966L...6L}.
Specifically, those TR loops have effective temperatures in a range of $1.9\times10^5$\,K--$1.3\times10^6$\,K in O\,{\sc iv}, by contrast to 
$2.2\times10^5$\,K--$2.3\times10^6$\,K in Si\,{\sc iv}\,\citep{2024ApJ...966L...6L}.
Those results imply that the heating energy effect on ions at $8\times10^4$\,K is stronger than that at $1.4\times10^5$\,K.

\par
The relations among parameters of loops are critical to infer the heating mechanisms therein\,\citep{1978ApJ...220..643R,2011ApJ...736....3V,2019ApJ...880...80B}.
The first study on these relations of TR loops\,\citep{2024ApJ...966L...6L} has revealed their extremely diverse natures, and the distribution may not be converged to a relation (see Figure\,\ref{fig07}).
One can also see that the goodness of the fitting obtained in Figure\,\ref{fig07} is modest.
This suggests that TR loops tend to have a great deal of variety.
Nevertheless, based on regression analysis, a reliable relation between electron densities ($n_e$) and loop length ($L$) is given as $$log_{10}(n_e)\varpropto -0.78 log_{10}(L).$$
This is much different from those found in coronal loops\,\citep[see e.g.][]{Aschwanden_1999,2014PASJ...66S...7B}.
For example, a scaling law of $log_{10}(n_e)\varpropto -0.41log_{10}(L)$ has been reported for isothermal coronal loops\,\citep{Aschwanden_1999}.
Owing to this relation, together with their diverse nature as revealed in Figure\,\ref{fig07}, TR loops must be heated in a very different way than their coronal counterparts.

Regarding the heating profiles of active region TR loops,
they are well believed to be heated impulsively\,\citep[e.g.][etc.]{2015ApJ...810...46H,2020ApJ...894..155S,2021NatAs...5..237B}.
With high resolution observations from SolO/EUI, a statistics of 42 brightenings in TR loops\,\citep{zuo2026} has been carried out.
The statistics shows that the heatings are impulsive and mostly located near the footpoints.
The impulsive phase of the brightenings has a time scale of $118.4\pm12.0$\,s, and the decaying phase of the brightenings has a time scale of $159.4\pm16.6$\,s.
Furthermore, impulsive heating drives enthalpy flows with a velocity of $51.3\pm5.6$\,\kms\ that is the most important mechanism to distribute the energy along the loop lengths.
The brightenings in TR loops could be explained by magnetic-reconnection-mediated impulsive heating at braiding sites of multiple loop strands\,\citep{2021NatAs...5..237B}.

\begin{figure}[!ht]
\centering
\includegraphics[width=\textwidth,clip,trim=0cm 26cm 0cm 0cm]{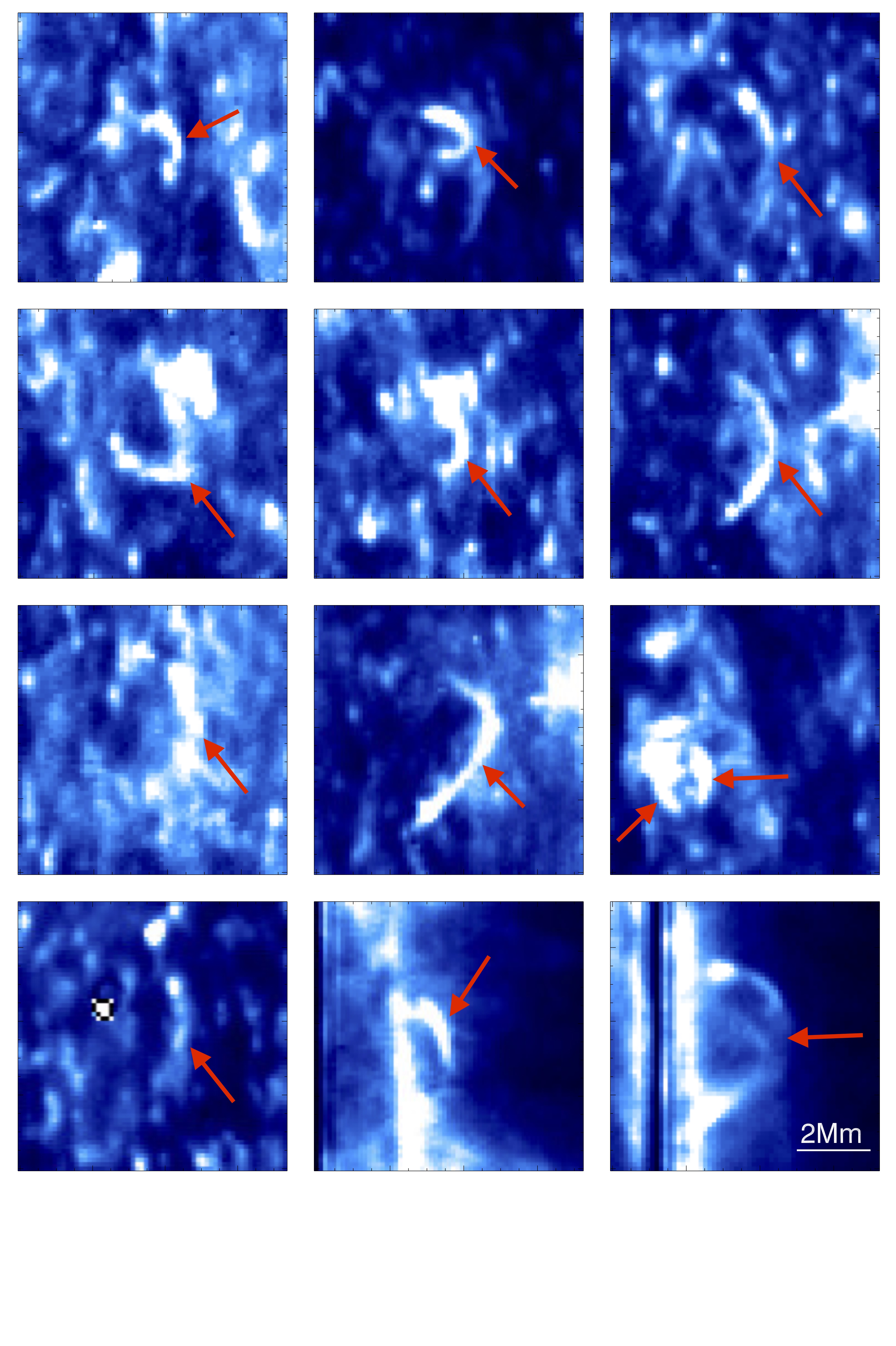}
\caption{Examples of TR loops in quiet sun region observed by IRIS slit jaw (SJ) 1400\,\AA\ images.
TR loops appear as arcade-like bright features denoted by red arrows.
The image scale is shown in the bottom-right panel.}
\label{fig08}
\end{figure}

\subsection{TR loops in quiet sun regions}\label{ssec32}
Magnetic loops in quiet sun regions are not rare, and a typical phenomenon is a coronal bright point (CBP).
A CBP has a size less than 50\,Mm that consists of multiple loops with various temperatures\,\citep{2019LRSP...16....2M}.
Although some loops in coronal bright points have cool legs with TR temperatures\,\citep{1981SoPh...69...77H,2008ApJ...681L.121T,2012ApJ...757..167K}, such a feature with TR temperature along its full length (i.e., TR loop) has hardly reported.
Based on spectral data from SUMER in O\,{\sc vi} ($T\sim2.8\times10^5$\,K), a loop-like feature along with supersonic flows in quiet sun region with a size of about 15\,Mm in length was reported\,\citep{2004A&A...427.1065T}, but convincing evidence of TR loops in quiet-sun regions is only available until that IRIS is in orbit\,\citep{2014Sci...346E.315H}, and some examples are shown in Figure\,\ref{fig08}.
Based on IRIS observations, TR loops in the quiet sun region have a full length of 4 to 12\,Mm and a maximum height of 1 to 4.5\,Mm\,\citep{2014Sci...346E.315H,2025ApJ...983..144B}.

\par
Small-scale loops (or loop-like features) in quiet sun region has also been observed by EUI aboard SolO in 174\,\AA\ passband with unprecedentedly-high resolutions, which have a similar size as the TR loops seen by IRIS\,\citep{2021A&A...656L...4B,2021A&A...656L..16M,2025A&A...699A.138N}.
Because the response function of EUI 174\,\AA\ passband has a large contribution from TR temperatures, some of these small-scale loops observed by EUI are likely TR loops\,\citep{2024A&A...688A..77D,2026arXiv260328601D}.
The interaction between such a loop and ambient field may lead to component reconnections and thus could be a good candidate for coronal heating mechanism in the quiet sun\,\citep{2021A&A...656L...7C}.

\par 
The TR loops in quiet sun region should be associated with weak magnetic field in the photosphere at granular scales\,\citep{2007A&A...475.1101S}.
The connection between TR loops and granular activities remains unclear. 
A comparison between IRIS observations and 3D models shows that these TR loops in quiet sun is also episodically heated, and the efficient radiative cooling due to high density prevents the occurrence of coronal temperatures in these structures\,\citep{2014Sci...346E.315H}.
However, both observational and theoretical studies on this group of loops are still very sparse.
Further investigations based on higher-resolution observations and/or more advanced numerical simulations are critical to better understand their natures and their roles in linking different layers of the solar atmosphere.

\begin{figure}[!ht]
\centering
\includegraphics[width=\textwidth]{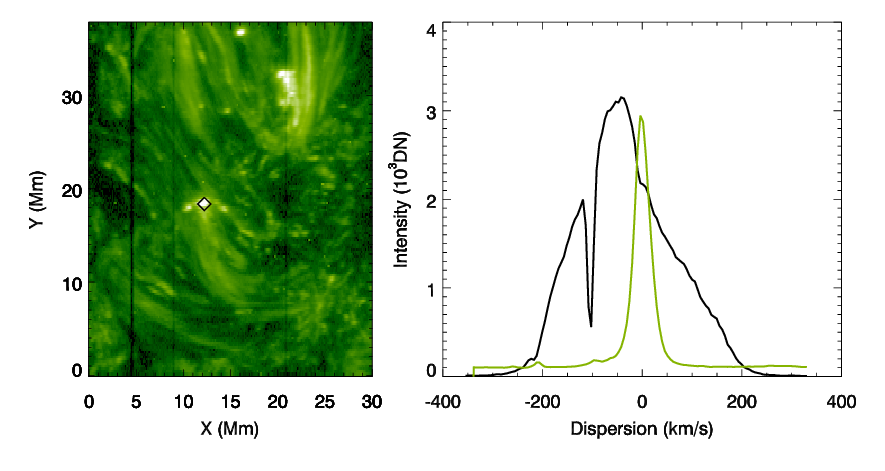}
\caption{Left panel: A region of the sun in IRIS Si\,{\sc iv} spectral raster, in which TR loops are abundant (bright threads in the image).
A UV burst is marked by a diamond symbol, an intense brightening in the image. 
Such bursts are rooted in numerous TR loops.
Right panel: The Si\,{\sc iv} 1394\,\AA\ spectral profile of a pixel of the UV burst (black line), compared to a reference profile that is obtained from averaging over a large region (green line).
The dip shown in the UV burst spectral profile at the dispersion of $-100$\,km/s results from the absorption of Ni\,{\sc ii}.} 
\label{fig09}
\end{figure}

\begin{figure}[!ht]
\centering
\includegraphics[width=\textwidth]{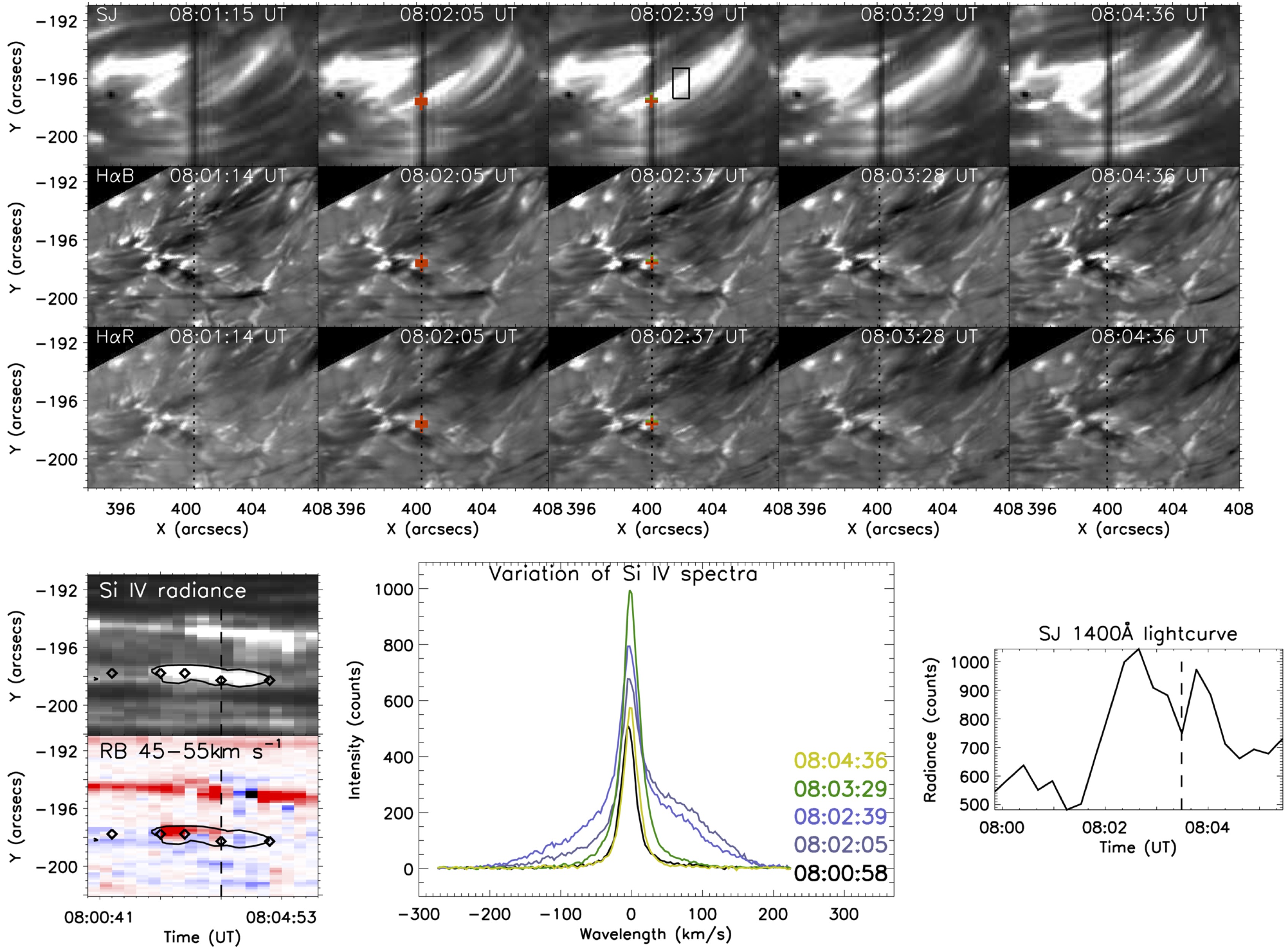}
\caption{Evolution of active TR loops observed by IRIS SJ images and their chromospheric response seen in the blue ($-35$\,\kms, H$\alpha$B) and red ($+35$\,\kms, H$\alpha$R) wings of H$\alpha$ by the Swedish Solar Telescope.
The variations in Si\,{\sc iv} radiance and RB asymmetry at $45-55$\,\kms, taken from, are shown in the bottom-left panel, which indicates that the activity has a lifetime of about 3 minutes.
The variations of Si\,{\sc iv} profiles from the location marked as red crosses in the top rows are shown in the bottom-middle panel, which shows non-Gaussian profiles during the activity (see examples as the curves at 08:02:05\,UT and 08:02:39\,UT).
The bottom-right panel shows an SJ 1400\,\AA\ light curve taken from the box region marked in the SJ image.
The figure is adapted from\,\citet{2017MNRAS.464.1753H}.} 
\label{fig10}
\end{figure}

\begin{figure}[!ht]
\centering
\includegraphics[width=\textwidth]{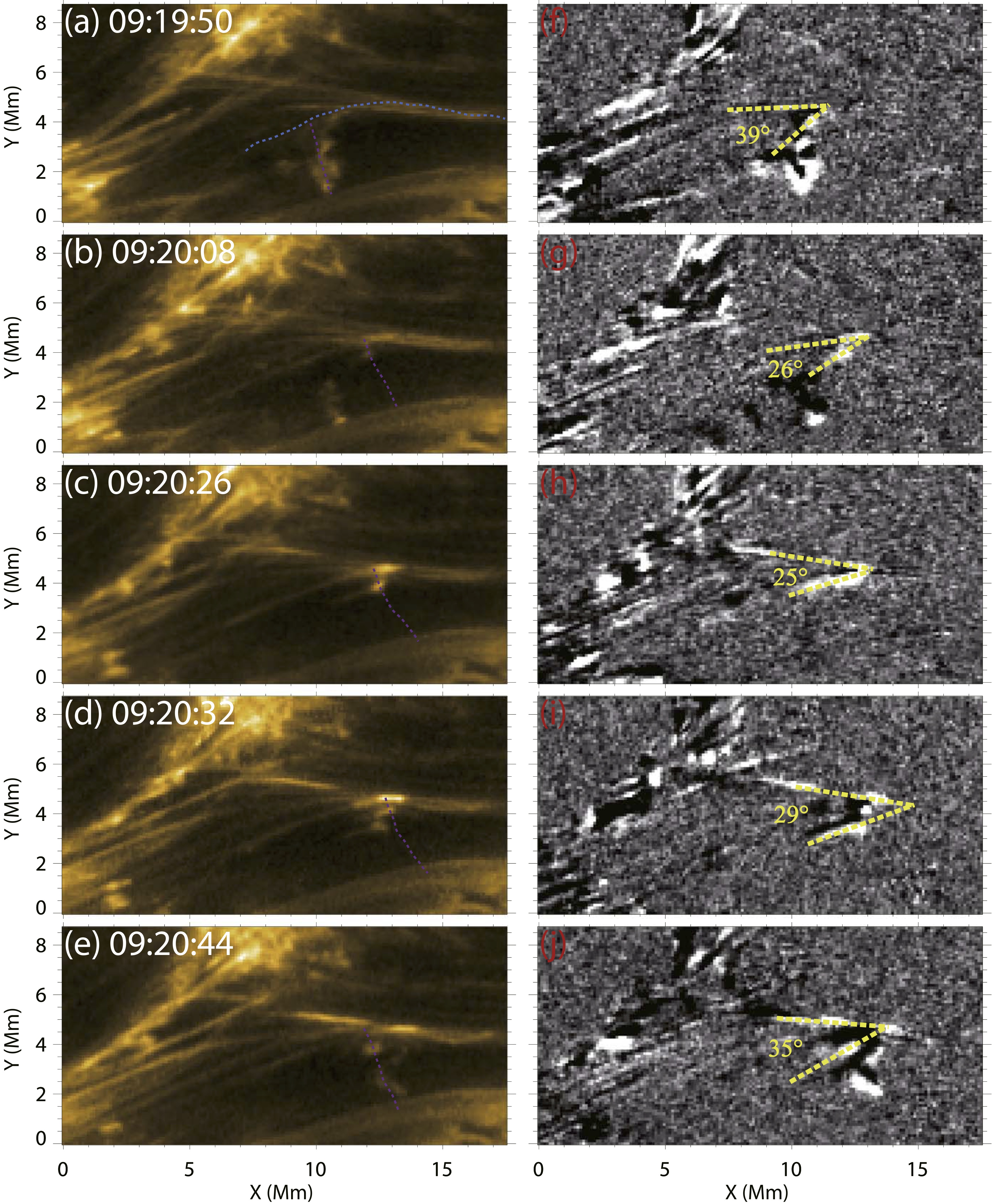}
\caption{Plasma ejections resulting from unbraiding of TR loops observed by SolO/EUI in 174\,\AA\ passband.
The left column shows the evolution of the region in the EUI original images, and the right column shows the same in the running-different images.
The main axis of the braiding of TR loops is denoted by the blue dashed line (panel (a)).
The purple dashed lines indicate the trajectory of the ejections of blobs due to the unbraiding of the TR loops.
The angles between the unbraided loop threads are shown in the right column.
The figure is adapted from\,\citet{2025ApJ...985...17Z}.} 
\label{fig11}
\end{figure}

\begin{figure}[!ht]
\centering
\includegraphics[width=\textwidth]{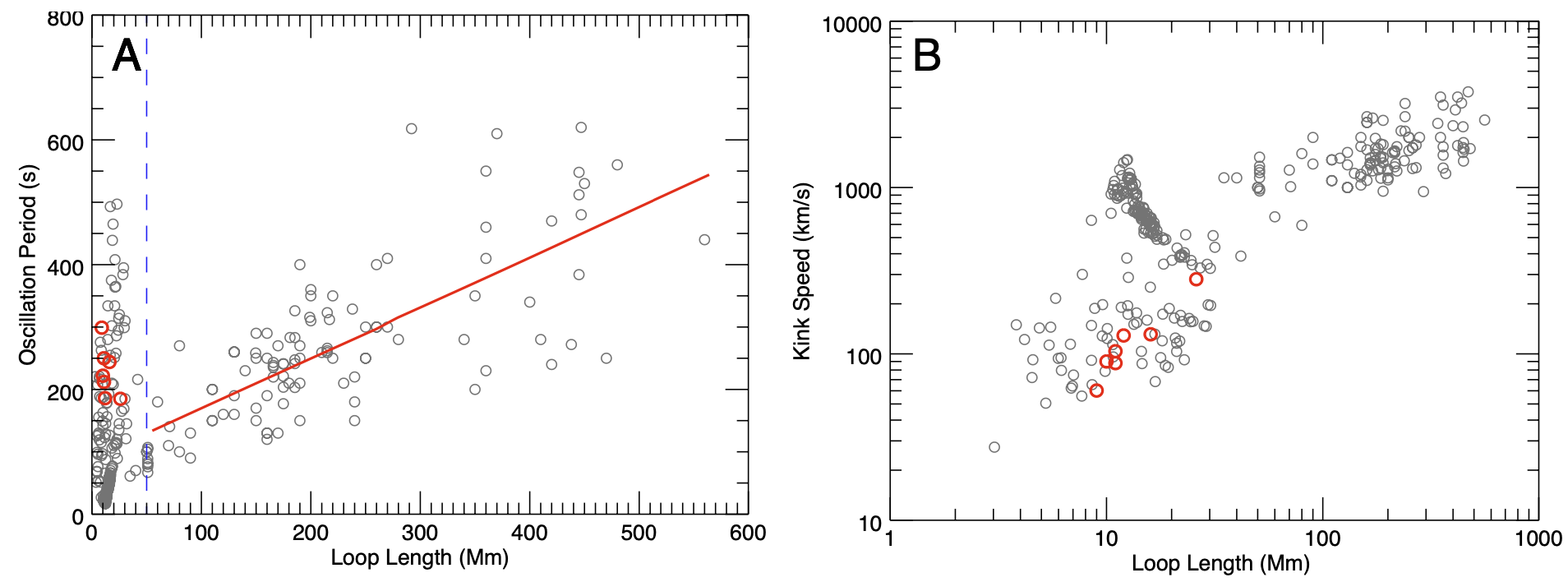}
\caption{Relationships between oscillation periods and loop lengths (panel A) and between kink speeds and loop lengths (panel B).
The decayless oscillations observed in TR loops are shown in red circles.
The figure is adapted from\,\citet{2024A&A...681L...4G}.} 
\label{fig12}
\end{figure}

\subsection{Small-scale activities associated with TR loops}\label{ssec33}
As described above, TR loops could be nascent magnetic structures in the solar atmosphere.
Therefore, activities in TR loops provide a good window into how the solar atmosphere is formed and shaped.
Multiple types of small-scale dynamics associated with TR loops have been discovered in the past decade since the launch of IRIS.

\par
UV bursts\,\citep{2018SSRv..214..120Y}, which are compact and intense radiation enhancements in TR spectra, frequently occur at the footpoints of TR loops (see Figure\,\ref{fig09}), especially at the joint footpoints of multiple loops\,\citep{2014Sci...346C.315P,2021Symm...13.1390H}.
This phenomenon is believed to be consequences of magnetic reconnections in the photosphere during flux emergence that heat the solar plasma from much lower temperatures to TR temperatures as demonstrated by both observations\,\citep{2014Sci...346C.315P,2017ApJ...836...63T,2018ApJ...854..174T,2021Symm...13.1390H} and numerical experiments\,\citep{2019A&A...628A...8P,2021A&A...646A..88N,2022A&A...665A.116N,2024A&A...685A...2C}, although whether the locations of the associated magnetic reconnection must be in the photosphere remains under debate\,\citep{2017ApJ...839...22H,2017RAA....17...31F,2019A&A...626A..33H,2020A&A...633A..58O}.

\par
Explosive events, which describe TR features emitting non-Gaussian line profiles with enhanced wings at Doppler velocities of 50--150\,\kms, are possible signatures of magnetic reconnections in the TR\,\citep{1989SoPh..123...41D, 1997Natur.386..811I}.
This phenomenon has been observed at the joint footpoints of multiple TR loops, with the interaction of opposite polarities underneath \citep{2015ApJ...810...46H}, suggesting magnetic reconnection between TR loops.
They are also found in the middle part of TR loop brightenings (see Figure\,\ref{fig10}) or associated with propagating bright blobs in TR loops\,\citep{2017MNRAS.464.1753H}, which might be due to splitting motions of or braiding reconnections among TR loop threads\,\citep{2021NatAs...5..237B}.

\par
Magnetic reconnection due to braiding of loops is an important process to heat the plasma in the solar atmosphere\,\citep{1983ApJ...264..635P,1983ApJ...264..642P,1988ApJ...330..474P}.
However, observations of magnetic braiding in TR loops are rare.
Such a study has recently been published based on observations from SolO\,\citep{2025ApJ...985...17Z}.
In those observations, magnetic braiding in TR loops generates a series of small-scale ejections of plasma blobs, which are constrained by post-reconnection field lines (see Figure\,\ref{fig11}).
The typical magnetic free energy producing a blob is estimated to be about $3.4\times10^{23}$\,erg, of which about $2.3\times10^{23}$\,erg goes into the kinetic energy, accelerating the plasma to about 90\,\kms.
The behaviour of magnetic braiding reconnection in TR loops is different from that occurring in the chromosphere\,\citep{2020ApJ...899...19C,2021NatAs...5...54A} or the corona\,\citep{2013Natur.493..501C,2022A&A...667A.166C}.

\par
Periodic phenomena in TR loops have also been reported, though such studies remain rare.
In emerging TR loops, quasi-periodic brightening at the loop apex with a period of about 3~min has been seen with IRIS data\,\citep{2019ApJ...887..221H}.
Decayless oscillations in TR loops with periods ranging from 3 to 5 min have also been reported, which have phase speeds that increase with loop length\,\citep{2024A&A...681L...4G}.
The oscillation behaviour of TR loops is similar to that of short loops, but different than that of the long coronal loops (Figure\,\ref{fig12}).
Such oscillations in TR loops provide a tool to probe the magnetic field strength in the region.

\section{Modelings of TR loops}\label{sec4}
A three-dimensional modeling based on the MURaM code\,\citep{2017ApJ...834...10R} has shown that TR loops as seen in Si\,{\sc iv} could be consequences of localised recurrent heating events driven by convective motions together with the complexity of magnetic features at the footpoints\,\citep{2023A&A...672A..47S}.
The magnetic reconnection, which is responsible for the heating events, might not require any flux emergence\,\citep{2023A&A...672A..47S}.
Regarding heating profiles in TR loops, a recent numerical experiment based on the hydrodynamics and radiative emission code \,\citep[HYDRAD,][]{2011ApJS..194...26B} shows that TR loops are heated impulsively with a complex heating profile that consists of multiple heating scales\,\citep{2024ApJ...971...59B}.
From simulations based on Bifrost code\,\citep{2016A&A...585A...4C} suggest that cold loops such as TR loops are a natural result of impulsive magnetic energy release by reconnection of magnetic braidings at low heights in the solar atmosphere\,\citep{2013ApJ...771...21W}.
The impulsive heating events in TR loops can then drive plasma flows therein\,\citep{2020ApJ...894..155S}.
These numerical experiments also emphasize that the non-equilibrium ionization in the TR is critical for the energy distribution during the heating of TR loops.

\par
For TR loops in the quiet sun, 3D simulations using Bifrost suggest that those short TR loops are consequences of low-lying magnetic structures coupling with plasma that is episodically heated\,\citep{2014Sci...346E.315H}. 
A study comparing SolO/EUI observations and simulations results from 1D hydrodynamics (HD) code HYDRAD again suggests that TR brightenings are results of TR loops in the quiet sun subjected to impulsive heatings\,\citep{2025A&A...699A..61D}.

\par
Although modelling studies specified on TR loops are sparse,
many simulations based on Bifrost have shown a great potential in resolving the TR features including loops\,\citep[e.g.][etc.]{2014Sci...346E.315H,2017ApJ...847...36M}.
Numerical experiments based on advanced modern codes like Bifrost and MURaM merit further investigation on the aspect of TR loops in the future.

\section{Summary and outlooks}\label{sec5}
TR loops are arcade-like features in the solar atmosphere observed at transition-region temperatures, representing the coupling between the under-heated, partially ionized plasma and the magnetic field in the TR.
They provide an important window for studying how energy and mass are transferred from the chromosphere to the corona through the TR.
They might form directly alongside flux emergence and thus help understand how the emerging magnetic flux reshapes the solar atmosphere.
Observations have also revealed that TR loops are very different than their coronal counterparts.
They are transient features in the solar atmosphere that are heated impulsively and are associated with many small-scale energetic events.

\par
Despite the growing focus on TR loops following the IRIS mission, studies on this subject, both observational and theoretical, are still very insufficient.
Some critical questions remain to be answered:\\
-- What is the upper limit of the sizes of TR loops? 
The upper limit of their sizes is crucial in understanding the coherent range of physical processes in the TR.
TR loops are better identified and measured with a spectrograph rather than imager because of the large temperature gradient in the region.
However, the operation of a normal spectrograph takes a long time to raster a large region and thus impossible to capture a large TR loop evolving rapidly.
Such a study will require spectral observations with larger field-of-view acquired in a very short time, which might be full-filled by the forthcoming Multi-slit Solar Explorer\,\citep[MUSE,][]{2020ApJ...888....3D}.\\
-- What are the properties of TR loops in the quiet sun?
This group of TR loops are still less known.
Their populations, sizes, heights, plasma flows, magnetic properties, and their connections to the lower and higher atmosphere are not fully understood.
To carry out such an investigation, higher resolution data from imagers, spectrographs, and magnetographs are required.\\
-- How do TR loops link to many other dynamic phenomena in the TR?
Since TR loops are representative of closed magnetic field lines in the region, how they link to the other dynamic phenomena is important to understand how the magnetic field governs the plasma evolution of the region.\\
-- How do heating events take place in TR loops?
A wide variety of dynamics in TR loops have been attributed to bald patch topology of emergence flux\,\citep{2017ApJ...836...63T}, twisting and braiding topology of the emerging flux\,\citep{2025ApJ...995...94C}, or interactions among different loops or ambient field\,\citep{zuo2026,2022ApJ...936...51W}.
How they are really driven, and their modulated mechanisms are not yet clear, and further investigation with high-resolution data and advanced simulations is warranted.\\
-- How does the energy released in TR loops transfer to the ambient atmosphere?
TR loops are very dynamic, but how their dynamics connect to the ambient atmosphere remains unknown.
This is crucial for determining the role of the TR in the entire chain of energy and mass transport in the solar atmosphere.\\
--  Whether and how are TR loops evolved to coronal loops? 
Heating events in TR loops are frequently observed, but so far, there is no observational evidence showing that they are heated to the corona or that they are cooled down.
This is due to the lack of observations with sufficient spatiotemporal resolutions and/or continuous representative temperatures, especially the latter one.
High-resolution data from the Daniel K. Inouye Solar Telescope\,\citep[DKIST,][]{2020SoPh..295..172R} and the forthcoming 2.5-meter Wide-field and High-resolution Solar Telescope (WeHoST) combined with those observed in more spectral lines from the forthcoming MUSE and Extreme Ultraviolet High-Throughput Spectroscopic Telescope (EUVST) should shed more light on the subject how TR loops are connected to the cooler and hotter layers of the solar atmosphere.\\


\bmhead{Acknowledgements}
The author is grateful to the anonymous referees for their constructive comments and suggestions, which make the review more clear and comprehensive.
The author also wants to thank Prof. Maria S. Madjarska and Prof. Bo Li for careful readings and comments on the first draft of the manuscript.
This research is supported by the Fundamental Research Funds for the Central Universities (KG202506), the National Natural Science Foundation of China (42230203, 42174201), the National Key R\&D Program of China (2021YFA0718600) and China's Space Origins Exploration Program.
Z.H. acknowledges the support by the International Space Science Institute (ISSI) in Bern, through ISSI International Team project \#24-605 (Small-scale magnetic flux ropes under the microscope with Parker Solar Probe and Solar Orbiter).
IRIS is a NASA small explorer mission developed and operated by LMSAL, with mission operations executed at NASA Ames Research Center, and major contributions to downlink communications funded by ESA and the Norwegian Space Centre.
The AIA data are used courtesy of NASA/SDO, the AIA and HMI teams, and JSOC.

\section*{Declarations}
The author declares there is no conflict of interest.


\bibliography{rmpp_revision}

\end{document}